\documentclass[twocolumn]{aastex631}

\usepackage{natbib}
\usepackage{epsfig}
\usepackage{graphicx}
\usepackage{subfigure}
\usepackage{float}
\usepackage{amsmath}
\usepackage{color}
\usepackage{amssymb}
\usepackage{apjfonts}

\newcommand{\UA}{Department of Physics, The Applied Math Program, and Department of Astronomy, The University of Arizona, Tucson, AZ 85721, USA}

\shortauthors{Melia}

\begin{document}

\title{A Pop III generated dust screen at $z\sim 16$}

\correspondingauthor{Fulvio Melia}
\email{fmelia@arizona.edu}

\author{Fulvio Melia \thanks{John Woodruff Simpson Fellow.}}
\affiliation{\UA}

\begin{abstract}

The search for alternative cosmological models is largely motivated by the
growing discordance between the predictions of $\Lambda$CDM and the ever
improving observations, such as the disparity in the value of $H_0$ measured
at low and high redshifts. One model, in particular, known as the $R_{\rm h}=ct$
universe, has been highly successful in mitigating or removing all of the
inconsistencies. In this picture, however, the anisotropies in the cosmic
microwave background (CMB) would have emerged at a redshift $z\sim 16$, rather
than via fluctuations in the recombination zone at $z\sim 1080$. We demonstrate
here that a CMB created in the early Universe, followed by scattering
through a Pop III generated dust screen, cannot yet be ruled out by the current
data. Indeed, the {\it Planck} measurements provide a hint of a $\sim 2-4\%$
frequency dependence in the CMB power spectrum, which would be naturally explained
as a variation in the optical depth through the dust, but not a Thomson
scattering-dominated recombination environment. Upcoming measurements should
be able to easily distinguish between these two scenarios, e.g., via the
detection of recombination lines at $z\sim 1080$, which would completely
eliminate the dust reprocessing idea.

\end{abstract}

\keywords{Cosmic background radiation (322) --- first stars (1285) --- reionization (1383) 
--- Cosmological parameters (339) --- Cosmological models (337) --- large-scale structure
of the Universe (902)}

\section{Introduction}\label{introduction}
Given how well we understand the origin of the cosmic microwave
background (CMB) and the underlying physics of the medium where it was 
produced, it might seem surprising to suggest that the anisotropies 
we observe in its temperature distribution perhaps emerged at a redshift 
of only $\sim 16$ rather than the conventional $z\sim 1080$. And yet, 
as we shall demonstrate in this paper, there are good reasons for
considering such a radical possibility. Even more to the point, we shall
show that none of the data available today clearly rule out a scenario
in which the CMB, though produced in the early Universe, was 
subsequently reprocessed by dust created in the ejecta of Pop III stars 
at the lower redshift. As revolutionary as this alternative picture may 
appear to be, an early version of it was actually seriously considered 
in the 1970's before the current paradigm was established. We aim to 
resurrect it here because there are growing indications that the tension 
developing between the predictions of $\Lambda$CDM and the ever improving 
observations may be completely resolved by an expansion history in which 
this Pop III dust screen scenario becomes viable.

We begin by motivating this proposal in \S\S~\ref{background} and
\ref{anisotropies}, and then proceed in \S~\ref{observations} to 
systematically examine why the current data do not yet rule out this 
picture.  We also provide some indication of how future, more precise 
observations, may be able to clearly distinguish between the recombination 
and dust-reprocessing models, e.g., via the detection of recombination 
lines in the CMB spectrum. We end with our conclusion in \S~\ref{conclusion}.

\section{Background}\label{background}
A pivotal development in our understanding of how the CMB was produced in the 
standard model occurred with COBE's discovery of a near perfect blackbody in
its spectrum \citep{Mather:1990}. All succeeding measurements of this relic 
signal have bolstered the view that the background radiation must therefore
have been thermalized within one year of the Big Bang, diffused through a
scattering-dominated medium, and eventually streamed freely after the protons 
and electrons combined \citep{Hinshaw:2003,Planck:2014}.

Prior to this period, however, the CMB was thought to have been produced by dust 
at lower redshifts, probably injected into the interstellar medium (ISM) by
Pop III stars \citep{Rees:1978,Rowan-Robinson:1979,Wright:1982}. In the absence 
of any indication to the contrary, it was also assumed that the radiation
rethermalized by this dust was itself emitted by the same stars.

But in the context of $\Lambda$CDM, there are several fundamental reasons why 
the CMB must have been produced at the epoch of recombination, $z_{\rm cmb}\sim
1080$. The dust-screen scenario we consider here cannot alter this situation. 
The surface of last scattering (LSS) had to lie at this redshift because 
(i) the characteristic scale observed
in the CMB's power spectrum, interpreted as an acoustic horizon, requires
a decoupling of the radiation and the baryonic fluid $\sim 380,000$ yrs after 
the Big Bang, corresponding to the aforementioned $z_{\rm cmb}\sim 1080$
\citep{Spergel:2003}; (ii) assuming that the CMB propagated freely after 
this time, its temperature must have scaled according to $T(z)\propto (1+z)$. 
One may therefore use the Saha equation to estimate the LSS temperature
($T_{\rm cmb}$), and hence the redshift, at which the ionization fraction 
dropped to $\lesssim 50\%$. Thus, using $T_{\rm cmb}\sim 3,000$ K and a CMB 
temperature today of $2.728$ K, one infers that $z_{\rm cmb}$ must have
been $\sim 1,100$, confirming the value implied by the acoustic horizon. 
On the other hand, if dust were somehow involved, the implied temperature
($T\lesssim 50$ K) would correspond to a redshift $z\lesssim 20$, 
which is clearly in conflict with the acoustic-scale interpretation of 
the CMB power spectrum. Subsequent work has provided even more reasons to
abandon the dust scenario, given that Pop III starlight scattered by 
the ejected dust could not produce the observed CMB spectrum \citep{Li:2003}.
This conclusion has been affirmed with more recent work 
showing that the original formalism used to calculate the thermalization 
of starlight by metallic needles is probably not correct, requiring a
re-evaluation of the absorption cross-section of these dust particles 
over a wide wavelength range \citep{Xiao:2020}. 

So the issue is not whether the conventional recombination picture for the
formation of the CMB needs to be modified in the standard model. Rather, the 
growing tension between the predictions of $\Lambda$CDM and the new observations 
is motivating us to consider alternative cosmologies, one of which---the 
$R_{\rm h}=ct$ universe \citep{MeliaShevchuk:2012,Melia:2020}---has been 
particularly successful in resolving these inconsistencies. As we shall see
beginning with \S~\ref{anisotropies}, the origin of the CMB would have been 
quite similar in these two models, but the interpretation of the temperature 
anisotropies in $R_{\rm h}=ct$ requires them to have emerged at $z\sim 16$, 
rather than the conventional $z_{\rm cmb}\sim 1080$. Ironically, the 
most likely origin of these anisotropies would thus be fluctuations in
a dust screen at that redshift, echoing the original (though now abandoned) 
scenario proposed for the standard model.

A well-known example of the standard model's inability to account for
all of the cosmological data arises from the inconsistency of the Hubble
constant measured at high and low redshifts. The analysis of the CMB 
observed with {\it Planck} \citep{PlanckVI:2020} indicates a value, 
$H_0=67.6\pm0.9$ km s$^{-1}$ Mpc$^{-1}$, lower than that measured locally, 
with a corresponding matter fluctuation amplitude, $\sigma_8$, higher than 
that derived from the Sunyaev-Zeldovich effect. In fact, the Hubble constant 
measured by {\it Planck} in the context of flat (i.e., $k=0$) $\Lambda$CDM 
disagrees by $\sim 4.4\sigma$ with that measured using type Ia supernovae, 
calibrated via the Cepheid distance ladder, which instead implies a value 
$H_0=74.03\pm1.42$ km $\rm s^{-1}$ $\rm Mpc^{-1}$ \citep{Riess:2018}. 

But these are not the only measurements that show evidence of discordance in 
the interpretation of $H_0$ within the standard model. Clustering, weak 
lensing, and baryon acoustic oscillations (BAO), have yielded results 
similar to the CMB \citep{Abbott:2018b}, in contrast 
to a local distance ladder calibrated with the tip of the red giant branch, which
implies yet a different value of $H_0$ \citep{Freedman:2020}. A broader discussion
of this tension may be found in \cite{Verde:2019}. Clearly, the $4.4\sigma$ 
disparity in the expansion rate of the Universe at low and high redshifts 
refutes the expansion history predicted by the standard model.

Yet it would not be unreasonable to argue that such tensions are due to 
astrophysical reasons rather than the cosmology itself. This is certainly 
a possibility and further scrutiny will help determine whether this position 
is tenable in the long run. But unfortunately, the inconsistencies extend well 
beyond merely the measurement of $H_0$, as described at length in \cite{Melia:2020}.
For example, it is now becoming quite evident that the measured angular
correlation function requires a cutoff in the primordial power spectrum 
(see \S~\ref{angular} below) that makes it impossible for slow-roll inflation 
to have seeded the fluctuations while simultaneously solving the temperature 
horizon problem \citep{MeliaLopez:2018,LiuMelia:2020}. A recent compilation 
of tests such as this, and a direct comparison between the standard model 
and $R_{\rm h}=ct$, may be found in Table~2 of \cite{Melia:2018e}.

The success of $R_{\rm h}=ct$ in resolving many of the observed inconsistencies 
in $\Lambda$CDM motivates us to probe the consequences of its expansion history 
more deeply, particularly with regard to the CMB. As we shall see shortly, a firm 
prediction of this model is that the CMB temperature anisotropies must have 
emerged at $z\sim 16$ if the acoustic scale inferred from the multi-peaked power 
spectrum coincides with the BAO scale measured at lower redshifts (see 
\S~\ref{anisotropies}). But this redshift is special for other reasons. It sits 
well within the period of Pop III star formation, just prior to the epoch of 
reionization, which began at $z_{\rm EoR} \sim 14-15$. This was a time when 
dust was likely being ejected into the ISM. 

If there is any validity to the $R_{\rm h}=ct$ cosmology, this overlap between
the redshift at which the CMB temperature anisotropies must have emerged, and
the era of Pop III star formation, cannot be a coincidence. So while our goal
in this paper is not to question the established recombination picture for the 
origin of the CMB in $\Lambda$CDM, we shall examine in detail how the (now dated) 
dust model may still be viable in the context of $R_{\rm h}=ct$. Indeed, we shall
demonstrate that, while recombination does not work in this model, the dust model 
is unavoidable. A principal difference between the original dust scenario and
that developed here, however, is that reprocessing of radiation emitted by the
Pop III stars plays no role in creating the CMB we observe
today. Instead, the background radiation would still have originated in the
early Universe, as it did in $\Lambda$CDM, but would have been reprocessed
through the dust screen prior to the epoch of reionization. Thus, a critical 
difference between these two cosmologies is that the CMB anisotropies would 
reflect the large-scale structure at $z\sim 16$ in $R_{\rm h}=ct$, while they
would be a consequence of the conventional fluctuation spectrum at $z\sim 1080$
in the context of $\Lambda$CDM.

\section{CMB anisotropies at $z\sim 16$}\label{anisotropies}
There are several independent observational indicators suggesting that the 
CMB anisotropies originated at $z_{\rm dust}\sim 16$ in the context of $R_{\rm h}=ct$. 
We discuss the two most prominent ones below.

\subsection{Equality of the acoustic and BAO scales}\label{acoustic}
The first of these is based on the conventional assumption that the acoustic horizon
responsible for the CMB multi-peaked power spectrum is equal to the BAO scale 
measured at much lower redshifts. 

{\it Planck} has identified a scale, $r_{\rm s}$, in the temperature 
and polarization power spectra \citep{Planck:2014}, corresponding to an angular 
size $\theta_{\rm s}=(0.596724\pm 0.00038)^\circ$. This is interpreted as
an acoustic horizon, dependent on how far sound waves could have travelled
in the comoving frame before matter and radiation decoupled. The specifics
of how this scale was established are not central to the argument made here,
other than the requirement that it remained constant in the comoving frame
once the radiation began to stream freely following recombination.

In parallel with the scale $\theta_{\rm s}$ seen by {\it Planck}, a peak has 
also been identified in the two-point correlation function of galaxies 
and the Ly-$\alpha$ forest. The growth of fluctuations in the matter 
density is still linear at the BAO scale, allowing one to model it with 
low-order perturbation theory 
\citep{Meiksin:1999,Seo:2005,Jeong:2006,Crocce:2006,Eisenstein:2007b,Nishimichi:2007,Matsubara:2008,Padmanabhan:2009,Taruya:2009,Seo:2010}.
The peak measured via large galaxy surveys is thought to be due to the
aforementioned acoustic scale.

In other words, one assumes that the acoustic horizon established at decoupling 
remains fixed in the comoving frame, re-appearing much later in the guise
of the BAO feature. Of course, the proper size of the BAO `ruler' is not
the same as the proper size of the acoustic scale. These lengths
may be identical in the comoving frame, but the acoustic scale
expands as the Universe evolves, at a rate consistent with the
expansion factor $a(t)$. The physical BAO scale is thus much bigger 
than the acoustic length at the LSS, given by the ratio
$a(t_{\rm BAO})/a(t_{\rm cmb})$, which depends critically on the 
cosmological model. As we shall see shortly, it is this ratio 
that eliminates any possibility of the CMB scale having been set
at recombination in $R_{\rm h}=ct$, implying instead that it must
have been established at a redshift much smaller than $\sim 1080$. 

The BAO peak positions can now be measured to better than $\sim 1\%$
accuracy with galaxies and $\sim 1.4-1.6\%$ with the Ly-$\alpha$
forest, thanks to the introduction of reconstruction techniques 
\citep{Eisenstein:2007a,Padmanabhan:2012} that enhance the 
quality of the correlation functions. The recent galaxy 
measurements at $z\lesssim 0.7$ \citep{Alam:2016}, in combination
with the Ly-$\alpha$ forest observation at $z=2.34$ 
\citep{Delubac:2015,Font-Ribera:2014}, indicate that the BAO
scale is $\sim 147$ Mpc in the case of flat $\Lambda$CDM, and 
$r_{\rm BAO}\sim 131\pm 4$ Mpc for $R_{\rm h}=ct$ \citep{Melia:2022c}. One should
emphasize that these measurements exclude the use of BAO observations
based on photometric clustering and the WiggleZ survey \citep{Blake:2011}, 
which have much larger errors. 

In the $R_{\rm h}=ct$ cosmology, the angular-diameter distance is simply given as
\citep{MeliaShevchuk:2012,Melia:2020}
\begin{equation}
d_A(z)=\frac{c}{H_{0}}\frac{1}{(1+z)}\ln(1+z)\;,\label{eq:dA}
\end{equation}
while the Hubble distance is
\begin{equation}
d_{\rm H}(z)={c\over H_0}{1\over(1+z)}\;.\label{eq:dH}
\end{equation}
One can therefore see that the ratio ${\cal D}(z)\equiv d_A(z)/d_H(z)$
has the form 
\begin{equation}
{\cal D}(z)=\ln(1+z)\;,\label{eq:Dfinal}
\end{equation}
which is free of any parameters.

If we now set the acoustic and BAO scales equal to each other in the
comoving frame, we find that
\begin{equation}
{\cal D}(z)=\ln(1+z)=\frac{r_{\rm BAO}}{R_{\rm h}(t_0)\,\theta_{\rm s}}\;,\label{eq:Dmeasured}
\end{equation}
where $R_{\rm h}(t_0)=c/H_0$ is the gravitational (or Hubble) radius today
\citep{Melia:2018b}. One therefore infers that the dust screen must lie at
\begin{equation}
z_{\rm dust}=16.05^{+2.4}_{-2.0}\quad({\rm acoustic=BAO})\;,\label{eq:dustscreen1}
\end{equation}
corresponding to a cosmic time $t_{\rm dust}\approx 849$ Myr in the
evolutionary history of the $R_{\rm h}=ct$ universe. In deriving this value, 
we have simply used the Hubble constant (i.e., $H_0=67.6\pm0.9$ km s$^{-1}$ Mpc$^{-1}$)
measured by {\it Planck}, though the actual optimization for $R_{\rm h}=ct$ may 
differ by several percentage points. The error in $z_{\rm dust}$ has been propagated 
from the uncertainties in $\theta_{\rm s}$, $r_{\rm BAO}$ and $H_0$. The second
indicator that confirms this redshift is based on an entirely different analysis 
of the CMB spectrum, which we discuss next.

\subsection{Angular size of the largest mode in the CMB fluctuations}\label{angular}
Several large-angle anomalies have been observed in the CMB fluctuation spectrum by 
every major satellite flown to study the microwave background since the 1990's
(COBE, \citealt{Mather:1990}; WMAP, \citealt{Hinshaw:2003}; and {\it Planck}, 
\citealt{Planck:2016a}). One of the most prominent among them is the lack of 
large-scale angular correlations in the temperature distribution, contrasting with 
basic inflationary theory, which is disfavoured by the measured angular correlation 
function at a confidence level exceeding $3\sigma$ \citep{Copi:2015}. 

An in-depth analysis of the latest {\it Planck} data, focusing on this particular
issue \citep{MeliaLopez:2018,Melia:2021b,Sanchis-Lozano:2022}, has shown that 
the paucity of large-angle correlations is best explained by the presence 
of a cutoff, 
\begin{equation}
k_{\rm min}=(3.14\pm0.36)\times 10^{-4}\;{\rm Mpc}^{-1}\;,\label{eq:kmin}
\end{equation}
in the primordial power spectrum, $P(k)$. The value of $k_{\rm min}$ 
can easily discriminate between inflationary (e.g., $\Lambda$CDM) and 
non-inflationary (e.g., $R_{\rm h}=ct$) models because the quasi-de 
Sitter expansion in the former would have stretched all the quantum 
fluctuations beyond the Hubble horizon, thereby producing strong 
correlations at all angles upon re-entry (i.e., $k_{\rm min}\approx 0$). 
Non-inflationary cosmologies, on the other hand, would have created a 
CMB power spectrum with wavelengths no bigger than the size of the 
Hubble horizon at the time the background radiation was produced. 

In addition to its impact on the angular correlation funtion, the 
existence of $k_{\rm min}\not=0$ also signals the time at which inflation
could have started. And the delayed initiation time implied by 
Equation~(\ref{eq:kmin}) makes it difficult to account for the origin of 
the primordial power spectrum while allowing the Universe to have expanded 
by a sufficient number of e-folds to overcome the horizon problem. This issue 
has been explored elsewhere 
\citep{Destri:2008,Scacco:2015,Santos:2018,Handley:2014,Ramirez:2012,Remmen:2014,LiuMelia:2020}, 
however, so we won't revist it here. Instead, our primary focus is on how 
$k_{\rm min}$ could represent a measurable angular scale for non-inflationary 
models. 

In $R_{\rm h}=ct$, the mode wavelengths grew at the same rate as the Hubble 
radius, both proportional to $a(t)$, so quantum fluctuations never crossed 
back and forth across the Hubble horizon \citep{Melia:2019b}. The mode with 
the smallest wavenumber in this cosmology, and hence the longest wavelength, 
\begin{equation}
\lambda_{\rm max}(t)\equiv \frac{2\pi\, a(t)}{k_{\rm min}}\;,\label{eq:lmax}
\end{equation}
corresponds to the first fluctuation to have emerged out of the 
Planck domain into the semi-classical Universe. Thus, the ratio 
$\lambda_{\rm max}(t)/R_{\rm h}(t)$, where $R_{\rm h}(t)=c/H(t)$ is the Hubble 
radius at time $t$, would have remained fixed throughout cosmic evolution.

One can see right away that this essential concept is well supported by the {\it Planck} 
data. In the $R_{\rm h}=ct$ cosmology, quantum fluctuations began to form at about 
the Planck time, $t_{\rm Pl}$, with a maximum wavelength 
\begin{equation}
\lambda_{\rm max}(t_{\rm Pl})=\eta\,2\pi R_{\rm h}(t_{\rm Pl})\;,\label{eq:lPlanck}
\end{equation}
where $\eta$ is a multiplicative factor $\sim O(1)$ \citep{Melia:2019b}.
The measured value of $k_{\rm min}$ (Eq.~\ref{eq:kmin}) confirms this 
basic theory by showing that the ratio $\lambda_{\rm max}(t)/R_{\rm h}(t)$
has not changed over the Universe's entire expansion since the Planck era,
corresponding to an increase in $a(t)$ by over 60 orders-of-magnitude.
Putting $t=t_0$ in Equation~(\ref{eq:lmax}), one finds 
that $\lambda_{\rm max}(t_0)\approx 2.0\times 10^4$ Mpc, which is to be
compared with $\eta 2\pi R_{\rm h}(t_0)\approx \eta 2\pi\times 4.5\times 10^3$ Mpc. 
These two lengths are equal as long as $\eta=2/3$. But it must be emphasized that
any expansion different from that in $R_{\rm h}=ct$ would have produced 
highly divergent values of $\lambda_{\rm max}(t_0)$ and $R_{\rm h}(t_0)$ 
today. In other words, the near equality of $\lambda_{\rm max}(t_0)$ and
$R_{\rm h}(t_0)$ would be extremely unlikely if it were random, so 
already we see that the basic premise underlying the origin of quantum 
fluctuations in $R_{\rm h}=ct$ is borne out by the {\it Planck} measurements. 

This outcome is highly relevant to our identification of where the anisotropies
in the CMB were produced because it provides an entirely new length scale 
(i.e., $\lambda_{\rm max}$) one may use to estimate $z_{\rm dust}$. The 
optimization that completely removes the angular correlation anomaly is 
given as $u_{\rm min}=4.34 \pm 0.50$, where $u_{\rm min}\equiv k_{\rm min}
r_{\rm dust}$, in terms of the comoving radius ($r_{\rm dust}$) to the dust 
screen \citep{MeliaLopez:2018,Sanchis-Lozano:2022}.  It is easy to see that 
one may therefore write
\begin{equation}
u_{\rm min}=\frac{1}{\eta}\frac{R_{\rm dust}}{R_{\rm h}(t_{\rm dust})}\;,\label{eq:umin}
\end{equation}
where $R_{\rm dust}$ is the proper distance to the dust screen and $R_{\rm h}(t_{\rm dust})$
is the corresponding Hubble radius at that time. Therefore,
\begin{equation}
\ln(1+z_{\rm dust})=\eta u_{\rm min}\;,\label{eq:lnzdust}
\end{equation}
which yields 
\begin{equation}
z_{\rm dust}=17.05^{+8}_{-5}\quad({\rm largest\;CMB\;mode})\;.\label{eq:dustscreen2}
\end{equation}

Equation~(\ref{eq:dustscreen2}) is to be compared with (\ref{eq:dustscreen1}), which 
provided an independent measurement of this redshift. Clearly, these two estimates of 
the dust screen's location are highly consistent with each other, even though
they follow from two entirely different length scales in the
CMB anisotropies: (i) the acoustic peak in the CMB power spectrum, and 
(ii) the (unrelated) size of the largest fluctuation mode. These results
affirm our conclusion that the CMB anisotropies emerged during the Pop III era 
when interpreted in the context of $R_{\rm h}=ct$.

\section{Impact of a dust screen at $z\sim 16$}\label{observations}
\subsection{Recombination lines}\label{rec.lines}
One of the most straightforward tests to distinguish between a model in which
the CMB propagated freely out of recombination at $z_{\rm cmb}\sim 1080$, versus 
one in which the background radiation subsequently diffused through a dust screen at 
$z_{\rm dust}\sim 16$, involves the search for recombination lines. These should be 
present at some level in the microwave spectrum if the standard scenario is correct,
but dust reprocessing would have completely wiped them out. Extensive simulations 
have already been carried out to predict the intensity of the lines in the 
standard model \citep{Rubino-Martin:2006,Rubino-Martin:2008}. They show that the 
impact of line emission on the angular power spectrum of the CMB is probably quite
small, $\sim 0.1\mu{\rm K}$--$0.3\mu{\rm K}$. Nevertheless, the recombination
lines may still be distinguished from other effects because of their peculiar 
frequency and angular dependence. Thus, if future high sensitivity experiments 
measure such deviations with narrow-band spectral observations, the dust-screen 
scenario we are proposing in this paper would almost certainly be ruled out.

\subsection{The dust-reprocessed CMB spectrum}\label{spectrum}
Within the standard model, the plasma near the last-scattering surface is 
dominated by hydrogen and helium ions and their electrons, so its opacity 
is heavily influenced by Thomson scattering. This process can dilute the Planck 
spectrum produced at large optical depths, though not its shape, as the 
CMB photons diffuse through a progressively thinning photosphere. In other 
words, though the CMB intensity may differ somewhat from a true Planck 
function, its `colour' temperature cannot change \citep{Melia:2009}.

In contrast to this relatively simple scenario, one must take into account the
fact that the efficiency of dust absorption, $Q_{\rm abs}$, is frequency dependent 
\citep{Wright:1982} within the hypothesized dust screen at $z_{\rm dust}\sim 16$ 
in the $R_{\rm h}=ct$ universe. As we shall see shortly, however, the reprocessed 
radiation field can still be quite simple and featureless if the physical conditions 
bring the dust particles into thermal equilibrium with the background light. To gauge 
its impact on the CMB's spectral shape, we begin by assuming a density 
$n_{\rm d}(\Omega,t)$ of thermalizers at a temperature $T_{\rm d}(\Omega,t)$, in the 
direction $\Omega\equiv(\theta,\phi)$. As is well known, $Q_{\rm abs}$ of the 
thermalizers depends on several factors, including composition, orientation, 
geometry and frequency.

In terms of the intensity $I(\nu,\Omega)$, $I/\nu^3$ is invariant, so if we assume 
Kirchoff's law with isotropic emission by each radiating surface along the line-of-sight, 
we may write the intensity observed at frequency $\nu_0$ in the direction $\Omega$ as
\begin{eqnarray}
I(\nu_0,\Omega)&=&\langle\sigma\rangle \frac{2h\nu_0^3}{c^2}\int_0^{t_0}\,
\frac{dV(t)}{dt}dt\;n_{\rm d}(\Omega,t)\times\nonumber\\
&\null&\frac{\langle Q_{\rm abs}(\nu[\nu_0,t])\rangle}{d_L(t)^2}
P(\nu[\nu_0,t],T_{\rm d}[\Omega,t])\,e^{-\tau(\nu_0,\Omega,t)}\,,\qquad\label{eq:Inu0}
\end{eqnarray}
where $\langle\sigma\rangle$ is the average cross section of the thermalizers,
$\langle Q_{\rm abs}\rangle$ is an average over the randomly oriented thermalizers
in the field of unpolarized radiation, $d_L$ is the luminosity distance, $dV$
is the comoving volume element, and
\begin{equation}
P(\nu,T)\equiv \frac{1}{\exp(h\nu/kT)-1}\label{eq:Planck}
\end{equation}
is the mean number of photons per mode. In writing $\langle \sigma Q_{\rm abs}\rangle\approx
\langle\sigma\rangle\langle Q_{\rm abs}\rangle$, we have assumed that one type of dust
is dominant. The blackbody intensity is correspondingly 
\begin{equation}
B(\nu,T)\equiv \frac{2h\nu^3}{c^2}P(\nu,T)\;.\label{eq:Bnu}
\end{equation}
In these expressions (and all that follow below), the quantity
\begin{equation}
\tau(\nu_0,\Omega,t)=\langle\sigma\rangle \int_t^{t_0} dt\;
\langle Q_{\rm abs}(\nu[\nu_0,t])\rangle\,n_{\rm d}(\Omega,t)\label{eq:tau}
\end{equation}
is the optical depth due to thermalizers along the line-of-sight between time 
$t$ and $t_0$.

For simplicity, we assume a scaling law
\begin{equation}
n_{\rm d}(\Omega,t)=n_{\rm d}(\Omega,0)(1+z)^\epsilon\;.\label{eq:nd}
\end{equation}
Then, for convenience, we recast these integrals in terms of the redshift $z$:
\begin{eqnarray}
I(\nu_0,\Omega)&=&\tau_0(\Omega)\frac{2h\nu_0^3}{c^2}\int_0^\infty\,dz^\prime\,
\frac{(1+z^\prime)^{\epsilon-1}}{c E(z^\prime)}\times\nonumber\\
&\null&\hskip-0.2in \langle Q_{\rm abs}(\nu_0[1+z^\prime])\rangle
P(\nu_0[1+z^\prime]\,T_{\rm d}[\Omega,z^\prime])\,e^{-\tau(\nu_0,\Omega,z^\prime)}\,,\label{eq:Inu0z}
\end{eqnarray}
with
\begin{equation}
\tau(\nu_0,\Omega,z)=\tau_0(\Omega)\, \int_0^z dz^\prime\;
\frac{(1+z^\prime)^{\epsilon-1}}{c E(z^\prime)}\,\langle Q_{\rm abs}(\nu_0[1+z^\prime])\rangle\;.\label{eq:tauz}
\end{equation}
Here, we have defined the quantities,
\begin{equation}
\tau_0(\Omega)\equiv \frac{c}{H_0}\,\langle\sigma\rangle\,n_{\rm d}(\Omega,0)\;,\label{eq:tau0}
\end{equation}
and
\begin{equation}
E(z)\equiv \frac{H(z)}{H_0}\;.\label{eq:Ez}
\end{equation}

But notice that
\begin{equation}
\frac{d}{dz}e^{-\tau(\nu_0,\Omega,z)}=-\tau_0(\Omega)
\frac{(1+z)^{\epsilon-1}}{c E(z)}\langle Q_{\rm abs}(\nu_0[1+z])\rangle
e^{-\tau(\nu_0,\Omega,z)}\;,\label{eq:dtaudz}
\end{equation}
and so Equation~(\ref{eq:Inu0z}) may be written in the form
\begin{equation}
I(\nu_0,\Omega)=-\frac{2h\nu_0^3}{c^2}\int_0^\infty\,dz^\prime\,
P(\nu_0[1+z^\prime],T_{\rm d}[\Omega,z^\prime])
\frac{d}{dz^\prime}e^{-\tau(\nu_0,\Omega,z^\prime)}\;.\label{eq:Inu0new}
\end{equation}
Integrating this expression by parts then yields 
\begin{eqnarray}
\hskip0.1in I(\nu_0,\Omega)&=&B(\nu_0,T_{\rm d}[0])+\frac{2h\nu_0^3}{c^2}\int_0^\infty\,dz^\prime\,
e^{-\tau(\nu_0,\Omega,z^\prime)}\times\nonumber\\
&\null&\qquad\frac{d}{dz^\prime} P(\nu_0[1+z^\prime],T_{\rm d}[\Omega,z^\prime])\;.\label{eq:Inu0final}
\end{eqnarray}

The dust-reprocessed CMB intensity, $I(\nu_0,\Omega)$, is therefore a true blackbody 
(the first term on the right-hand side of Eq.~\ref{eq:Inu0final}), unless the second 
term---arising from modifications to the background radiation as it diffuses through 
the dust---is significant compared to the Planck function $B(\nu_0,T_{\rm d}[0])$. 
Very critically, however, it is not difficult to see that $I(\nu_0,\Omega)$ does not 
deviate at all from the Planck function when $T_{\rm d}(z)\propto (1+z)$, {\sl regardless 
of how the optical depth $\tau(\nu_0,\Omega,z)$ varies with $\nu_0$}.

To fully understand this point, consider that the temperature of free-streaming
radiation scales simply as
\begin{equation}
T(z) = T_0(1+z)\;.\label{eq:Tz}
\end{equation}
Thus, if the dust and the radiation it rethermalizes are in equilibrium (we shall
discuss what is required for this to happen in \S~\ref{temperature} below), 
$T_{\rm d}$ is also expected to follow Equation~(\ref{eq:Tz}). But the frequency 
itself scales as $\nu\propto (1+z)$, and so $P(\nu,T)$ is independent of redshift. 
The term $(d/dz^\prime)P$ in Equation~(\ref{eq:Inu0final}) is thus strictly zero,
leaving $I(\nu_0,\Omega)=B(\nu_0,T_{\rm d}[0])$ at each and every frequency 
(see \citealt{Rowan-Robinson:1979} for the introduction of such ideas).

\subsection{Temperature of the dust screen}\label{temperature}
We thus see that our intuition concerning the impact of dust reprocessing based 
on our local, galactic environment, may not apply to a dust screen at $z_{\rm dust}
\sim 16$. The key issue is---not that the dust opacity is frequency dependent but, 
instead---whether the physical conditions at this redshift may have induced the 
dust to reach local thermal equilibrium with the background radiation. If the 
answer is yes then, according to Equation~(\ref{eq:Inu0final}), the frequency 
dependence of the dust opacity is irrelevant to the final reprocessed radiation 
spectrum. 

In this section, we shall first consider whether Pop III stars could have generated 
enough dust particles to produce an optical depth $\tau(\nu_0,\Omega,z)\gg 1$ at 
$z_{\rm dust}\sim 16$, and then see if the condition $T_{\rm d}=T$ could have realistically
been met. In preparation for this analysis, we should remind ourselves that the
early work on this possibility, though framed in the context of $\Lambda$CDM, already 
demonstrated that a medium could be rendered optically thick just by dust, even if 
the latter represented a mere percentage density compared to other constituents in 
the cosmic fluid \citep{Rees:1978,Rowan-Robinson:1979,Rana:1981,Wright:1982,Hawkins:1988}.

A remnant trace of Pop III stars seeded prior to $z\sim 15$ appears to have been found
in the guise of extremely metal-poor stars in the galactic bulge \citep{Howes:2015},
consistent with the conventional picture of a very low metal abundance in the 
interstellar medium prior to stellar nucleosynthesis during the Pop III era. The 
actual metallicity between Pop III and Pop II star formation has not yet been identified,
however, so we parameterize it as $f_{\rm Z}$ relative to solar abundance. As we shall 
see shortly, the dust was created prior to $z\sim 16$ and then subsequently destroyed 
by Pop II supernovae, coinciding with the beginning of the epoch of reionization 
(at $z_{\rm EoR}\sim 14-15$). It is the dust's existence within this relatively 
narrow range of redshifts that motivates us to refer to it as a `dust screen.' 

The expansion profile in the $R_{\rm h}=ct$ universe is similar, though not identical,
to that in $\Lambda$CDM, so the optimized cosmological parameters, such as the Hubble
constant, can differ by several percentage points between these two models. Nevertheless,
for simplicity let us assume the concordance values for the most essential variables,
notably $H_0=67.7$ km s$^{-1}$ Mpc$^{-1}$ and a baryon fraction 
$\Omega_{\rm b}\sim 0.04$ \citep{Planck:2016a}. With these quantities, we estimate
that the comoving metal mass density at $z=16$ was $\rho_{\rm s}(z=16)\sim 4
\times 10^{-29}f_{\rm Z}$ g cm$^{-3}$. Thus, with a bulk density of silicate grains
$\sim 2$ g cm$^{-3}$, each of which had an average radius $r_{\rm s}\sim 0.1$ micron, 
the dust number density would have been $n_{\rm s}(z=16)\sim 5\times 10^{-15}f_{\rm Z}$
cm$^{-3}$.

We can thus estimate the metallicity $f_{\rm Z}$ required for all the CMB photons
to have been absorbed at least once by the dust screen by noting that, at $z=16$, 
its spectrum ranged from $\lambda_{\rm min}\sim 0.003$ cm to $\lambda_{\rm max}
\sim 0.02$ cm, corresponding to a dust absorption efficiency $Q(\lambda_{\rm min})
\sim 0.02$ and $Q(\lambda_{\rm max})\sim 0.003$ \citep{Draine:2011}. Thus, the 
photon mean free path due to dust absorption would have fallen in the range 
$3\times 10^{25} f_{\rm Z}^{-1}$ cm $\lesssim\langle l_\gamma\rangle\lesssim$ 
$2\times 10^{26} f_{\rm Z}^{-1}$ cm. This is to be compared with the gravitational 
(or Hubble) radius at that redshift, $R_{\rm h}(z=16)\sim 10^{27}$ cm \citep{Melia:2018b}. 
We therefore see that every CMB photon would have been absorbed by dust prior 
to $z\sim 16$ if $f_{\rm Z}\gtrsim 0.2$, representing $\sim 20\%$ of the solar 
value---a very reasonable number indeed. Moreover, a more recent
investigation \citep{Huang:2021} has shown that the absorption efficiency may
be even larger---by a factor of up to 10---compared to that reported by
\cite{Draine:2011}, so the required metallicity at $z\sim 16$ is probably
even smaller than $20\%$ of the solar value. 

The second issue concerns whether this absorption process was sufficient to
bring the dust particles into thermal equilibrium with the radiation, for 
which we must consider two additional factors. First, the average cooling 
$K(T_{\rm d})$ and heating $H(T)$ rates determine the energy flow to 
and from the radiation. Second, one must take into account the fact 
that each photon absorption may have produced a jump in the dust particle's 
temperature, depending on its size \citep{Weingartner:2001,Draine:2001}. 
The dust was heated by an isotropic radiation field with an angle-averaged
intensity $J_\lambda=B(\lambda,T)$ (see Eq.~\ref{eq:Bnu}), with $T(z=16)\approx 
46$ K in the cosmological context. Note that this differs from our local 
galactic neighborhood, where the primary heating agent is instead UV light.
In this environment, a typical dust particle was heated at a rate $H(T)=4\pi r_{\rm s}^2
\int_0^\infty d\lambda\,\pi B(\lambda,T)Q(\lambda)$, calculated from the previously
defined absorption efficiency $Q(\lambda)$. Kirchoff's law then gives its
emissivity $\propto B(\lambda,T_{\rm d})Q(\lambda)$. Similarly, its cooling 
rate was $K(T_{\rm d})=4\pi r_{\rm s}^2 \int_0^\infty d\lambda\,\pi 
B(\lambda,T_{\rm d})Q(\lambda)$. These two integrals are identical only
when $T_{\rm d}=T$.

The dust temperature evolves according to the equation
\begin{equation}
C(T_{\rm d}) \;dT_{\rm d}/dt=H(T)-K(T_{\rm d})\;,\label{eq:dTdt}
\end{equation}
where $C(T_{\rm d})$ is the heat capacity. At $T_{\rm d}\sim 46$ K, we have 
approximately $C(46\;{\rm K})\sim 0.2\, k_{\rm B}N_{\rm s}$, where $k_{\rm B}$ is 
Boltzmann's constant and $N_{\rm s}$ is the total number of atoms in 
the dust grain \citep{Draine:2001}. For the aforementioned grain size
$r_{\rm s}\sim 0.1$ micron, one estimates that $N_{\rm s}\sim 3\times 10^8$ 
\citep{Weingartner:2001}, and with $\langle Q(\lambda)\rangle\sim 0.012$, one 
therefore finds that 
\begin{equation}
dT_{\rm d}/dt\sim 10^{-7}(T^4-T_{\rm d}^4)\;{{\rm K}\,{\rm s}^{-1}}\;.\label{eq:dTdtnum}
\end{equation}
Let us assume that either $H(T)$ or $K(T_{\rm d})$ was dominant when $T_{\rm d}\not=T$.
We then see that it would have taken a mere 50 seconds for the dust to reach
equilibrium with the background CMB field at $T= T_{\rm d}\sim 46$ K. Subject
to possible limitations from the second factor we shall consider shortly, this
result thus clearly shows that the dust at $z\sim 16$ would have been fully
thermalized with the background radiation.

There is a caveat to this result, however, which constitutes the second factor 
we must take into account. This outcome would be unrealistic if the dust grains were 
so small that each absorption event would have changed their temperature in quantum 
jumps, rather than continuously. The absorption of a photon with wavelength $\lambda$ 
changes the temperature of a dust grain containing $N_{\rm s}$ atoms by
an amount $\Delta T_{\rm d}=hc/\lambda\,C(T_{\rm d})\sim 7.2\;(\lambda\,N_{\rm s})^{-1}$ 
K \citep{Weingartner:2001,Draine:2001}. Larger grains (i.e., those with radius
$r_{\rm s}\sim 0.1-0.3$ $\mu$m) would thus have avoided this problem because they 
contain $N_{\rm s}\sim 3\times 10^8-10^{10}$ atoms, for which $\Delta T_{\rm d}$
is a minuscule fraction ($\sim 10^{-9}-10^{-8}$) of $T_{\rm d}=46$ K at all the 
wavelengths of interest, $\lambda\sim 0.003-0.02$ cm. For them, the evolution in 
$T_{\rm d}$ at $z\sim 16$ would have proceeded smoothly, as described above.
But smaller grains have less heat capacity and a reduced radiating area, so 
each CMB photon absorption would have produced spikes in temperature 
\citep{Draine:2001}. 

The assumption of a smooth evolution in $T_{\rm d}$ thus breaks down for grains 
smaller than $r_{\rm s}\sim 0.003$ $\mu$m, for which thermal equilibration
would have proceeded via stochastic heating. So in summary, the dust temperature
in this model could very easily have been equilibrated to the background radiation
temperature as long as the dust particles were silicates of size $\sim 0.003-0.3$ $\mu$m.
Larger grains would have violated the previous estimate of $n_{\rm s}(z=16)$ and
$f_{\rm Z}\sim 0.2$, requiring an unreasonably large fraction of the mass in the
form of dust at $z\sim 16$.

Interestingly, such ultrasmall grains are destroyed faster than their bigger 
counterparts when exposed to the shock waves produced by the supernovae of Pop III 
stars (see \S~\ref{power.spec}), so there may be a natural motivation for expecting
an absence of these ultrasmall particles during the principal time when the CMB 
was rethermalized.

To complete our self-consistency check, we need to demonstrate that these estimates
do not violate our assumption of a negligible contribution to the overall background
radiation field by the Pop III stars. These stars were much more massive 
($500\;M_\odot\gtrsim M\gtrsim 21\;M_\odot$) than those forming today 
\citep{Bromm:2004,Glover:2005}, and emitted copious high-energy radiation 
that ionized the halos within which they grew \citep{Johnson:2007}. We can
estimate the fluence of radiation they emitted by calibrating it to how much 
metallicity they contributed to the interstellar medium. 

A large fraction of the Pop III stars exploded as SNe following their brief
($\sim 10^6-10^7$ yr) lives \citep{Heger:2003}, creating the dust screen
\citep{Whalen:2008} we are proposing in this paper. From the dust particle
size and number inferred above, we estimate that these stars must have
injected roughly $9\times 10^{44}$ g Mpc$^{-3}$ (co-moving volume) of dust 
into the interstellar medium during their principal epoch ($20\gtrsim z\gtrsim
15$) of formation.

Whether a Pop III star ended its life as a SN depended on its pre-explosion
mass. Its mass also provided an indication of how much metallicity it created
prior to ending its life: for $M\lesssim 40\;M_\odot$, it ejected $\sim20\%$
of its mass into the interstellar medium; for $M\gtrsim 140\;M_\odot$, the 
explosion was much more powerful, dispersing $\gtrsim 50\%$ of its mass 
\citep{Heger:2002}. For simplicity, we adopt a typical mass $M\sim 100\;M_\odot$
and an ejection fraction of $30\%$ (representing an average between these two 
limits). In the $R_{\rm h}=ct$ universe, 
\begin{equation}
1+z=1/tH_0\;,\label{eq:Ht}
\end{equation}
implying an interval of time $\Delta t \sim 200$ Myr between $z=15$ and $20$. 
And so we estimate that, during the principal Pop III era, an amount of dust 
required to render the interstellar medium optically thick could have been produced 
if $\sim 1.5\times 10^8$ Mpc$^{-3}$ of the stars exploded as SNe. 

A typical Pop III star with mass $M\sim 100\;M_\odot$ also emitted radiation
as a blackbody with radius $R_*=3.9\,R_\odot$ and a surface effective temperature 
$T_*=10^5$ K, implying a bolometric luminosity $\sim 4\times 10^{39}$ erg
s$^{-1}$. The total stellar energy density radiated during the Pop III era
would thus have been $U_{III}\sim 4\times 10^{63}$ erg Mpc$^{-3}$. At $z\sim 16$,
however, the CMB energy density was $U_{\rm cmb}\sim 8\times 10^{65}$ erg Mpc$^{-3}$.  
Therefore, $U_{III}/U_{\rm cmb}\sim 0.5\%$---a negligible fraction that would have
easily been thermalized and absorbed into the background Planck distribution. And 
if we consider photon number density instead of radiative energy density,
this ratio would have been even smaller, since the average stellar photon
energy was much higher than that of its CMB counterpart. 

\subsection{Frequency-dependent power spectrum}\label{power.spec}
A less discussed, though equally important, consequence of a dust screen for the
CMB is that, unlike the frequency-independent optical depth in a Thomson scattering 
environment, the depth to which our instruments `see' the radiation through the dust 
screen does change with frequency. This could result in a measurably different 
fluctuation pattern as a function of wavelength. 

Such photospheric depth effects would probably not change the shape and size of 
the larger fluctuations observed at different frequencies, but they could alter
the anisotropy observable on (smaller) scales comparable to the angular diameter
displacement of the dust screen at two different frequencies. At some level,
these differences would alter the CMB power spectrum constructed at one wavelength 
compared to that at another. 

\begin{table*}
\vskip 0.1in
\center
\centerline{{\bf Table 1.} Width of the dust screen at $z_{\rm cmb}\sim 16$
and its maximum impact on $\theta_{\rm s}$}
\begin{tabular}{cccc}
\hline\hline
$\Delta z$&{$\Delta t$}\qquad&{$\Delta\theta_{\rm s}$}\qquad&\qquad 
Percentage\qquad \\
 &{(Myr)}\qquad&{(deg)}\qquad&\qquad of $\theta_{\rm s}$ \\
\hline
1 & \; 53 & 0.013 &\qquad 2.2$\%$ \\
2 & 100 & 0.025 &\qquad 4.2$\%$ \\
\hline\hline
\end{tabular}
\end{table*}

Before we consider the expected differences in this effect between the recombination
and dust-reprocessing scenarios, let us first gather the observational evidence 
concerning a possible frequency dependence of the power spectrum. Such a 
detailed analysis was reported most recently by the {\it Planck} collaboration 
\citep{Planck:2016a}, following an earlier assessment of the WMAP first-year release 
(see, e.g., fig.~2 in \citealt{Hinshaw:2003}).

Generating the {\it Planck} maps at different frequencies is challenging due to
the wavelength dependence of the foreground conditions themselves, making it 
difficult to cross-correlate the patterns. In spite of this, {\it Planck} 
has shown that residuals in the half-mission TT power spectra, sampling a
frequency range $70-217$ GHz, clearly vary from one power spectrum to the 
next. The caveat here, of course, is that this variation could merely be 
due to non-cosmological factors, such as foreground systematics. 

On a more technical level, one may also estimate the dependence of the 
multipole power coefficients on the chosen frequency by extending the 
range of multipole numbers used in the analysis, up to a 
maximum value $\ell_{\rm max}$. Allowing $\ell_{\rm max}$ to vary from 
$\sim 900$ to several thousand permits one to search for a greater 
variation in the observed anisotropies on small scales compared to the 
larger ones.  Interestingly, this type of analysis shifts the optimized 
cosmological parameters by up to $\sim 1\sigma$, whose interpretation 
cannot always be made in terms of non-cosmological effects. Perhaps even 
more importantly, the cross power spectrum at lower frequencies ($\lesssim 
100$ GHz) reveals variations in the overall amplitude $D_{\ell}$ by as much
as $\sim 4\sigma$ compared to that inferred from measurements at higher
frequencies.

As of today, the {\it Planck} analysis reveals that the CMB multipole
power varies by an amount $\Delta D_\ell$ $\sim 40\;\mu$K$^2$ at $\ell\sim 400$, 
all the way up to $\sim 100\;\mu$K$^2$ at $\ell\gtrsim 800$. Aside from the fact 
that these variations appear to be real, it should also be noted that $\Delta 
D_\ell$ {\it increases} with multipole number $\ell$ across the frequency range 
$\sim 70-200$ GHz, as one would expect if a variable optical depth were
preferentially affecting the smaller fluctuations. Given that $D_\ell\sim 2000
\;\mu$K$^2$ over this range, the maximum variation of the power spectrum due to 
a frequency-induced change in the optical depth in the emission region appears 
to be $\sim 2\%$ at $\ell\sim 400$, increasing to $\sim 5\%$ at $\ell\gtrsim 800$. 
In other words, based on what we know so far, the frequency dependence of the
power spectrum is modest, but not absent. {\it We therefore conclude that the 
evidence does not appear to clearly favour a CMB produced within a purely Thomson 
scattering environment.}

This frequency dependence of the power spectrum instead appears to be qualitatively
consistent with what one would expect in the dust-screen scenario we are 
exploring in this paper. Let us also see if the current limits quoted above
stand up to more quantitative scrutiny. In order for it to be consistent with
the observations, the angular diameter distance to the dust screen should
not vary with frequency so much that it causes unacceptably large variations
in the inferred comoving BAO scale at $z_{\rm dust}\sim 16$. As discussed in 
\S~\ref{anisotropies} above, any variation in the optical depth through the 
dust screen would be limited to the range of angular diameter distances 
between $z\sim 14$ and $16$, which we shall justify more fully in this section.

Though nucleosynthesis and mass ejection in Pop III stars were somewhat 
different than those occurring later, our current understanding of the 
life-cycle of dust suggests it was quite similar then and now: 
(i) dust particles formed primarily in the ejecta of evolved stars; and 
(ii) were subsequently destroyed much more rapidly than they were formed 
in supernova-generated shock waves. This story line was established over
half a century ago \citep{Routly:1952,Cowie:1978,Seab:1983,Welty:2002}
with the earliest observations of shock-induced dust destruction, which
severely constrains how much dust can possibly exist near young, 
star-forming regions. The early-type stars emit most of the UV light
and evolve on a time scale of only $10-20$ Myr, ending their lives 
as supernovae. Their shocks completely destroy all the grains in the ISM
on a time scale $\lesssim 100$ Myr \citep{Jones:1994,Jones:1996}.

If we adopt the view that the dust screen was created by Pop III stars
prior to $z_{\rm dust}\sim 16$, we can then estimate the redshift, $z_{\rm dust}-\Delta z$, 
by which subsequent Pop III and Pop II supernovae would have completely destroyed
it. According to Equation~(\ref{eq:Ht}), $100$ Myrs at $z_{\rm dust}$ corresponds
to $\Delta z\sim 2.0$, while $\Delta z\sim 1.0$ is roughly $53$ Myrs. The
dust destruction scenario thus corresponds very closely to the other
observational requirements we have discussed thus far, which see the creation
of the dust screen due to Pop III stars by $z_{\rm dust}\sim 16$ and the beginning 
of the EoR by $z_{\rm EoR}\sim 14-15$.

Then, using Equation~(\ref{eq:Dmeasured}) with $z=z_{\rm dust}$, we can easily 
estimate the impact of such a dust-screen width on the inferred angular scale 
$\theta_{\rm s}$. Assuming the medium was optically thick at $z_{\rm dust}\sim 16$ 
and that it became transparent by $z_{\rm dust}-\Delta z$, one can show that 
$\theta_{\rm s}$ would have changed by an amount
\begin{equation}
\Delta\theta_{\rm s}=\frac{r_{\rm BAO}}{R_{\rm h}(t_0)}
\left[{\frac{1}{\ln(1+z_{\rm dust}-\Delta z)}-\frac{1}
{\ln(1+z_{\rm dust})}}\right]\;.\label{eq:delta.theta}
\end{equation}
The range of possible outcomes is shown in Table~1. Since the angular diameter
distance in the $R_{\rm h}=ct$ universe changes very slowly with redshift
near $z_{\rm dust}$, the maximum possible impact on $\theta_{\rm s}$ of optical 
depth effects across the dust screen is quite small, no more than a few 
percent---fully consistent with the current observational limits discussed above.

\subsection{E-mode and B-mode polarization}\label{polarization}
We next consider what the detection (or non-detection) of E-mode and B-mode
polarization can reveal about the medium where the CMB is produced and (possibly)
reprocessed by dust. We focus on linear polarization, which has already been
measured at some level, and promises to probe inflationary physics
at an unprecedented level of detail via several upcoming missions,
e.g., LiteBIRD \citep{Hazumi:2019}; COrE \citep{Delabrouille:2018}; PRISM 
\citep{Andre:2014}; PICO \citep{Hanany:2019}; and, more recently, ECHO 
\citep{Adak:2022}.

Linear polarization is geometrically decomposed into two rotational invariants,
known as the E (gradient) mode and the B (curl) mode \citep{Kamionkowski:1997,Zaldarriaga:1997}.
The Thomson scattering of anisotropic radiation associated with the scalar density 
fluctuations responsible for the temperature hot spots also creates E-mode polarization. 
This process does not produce B-mode polarization, however, because the modes are 
longitudinal density compressions aligned perpendicular to the direction of propagation, 
resulting in a pattern with zero curl. In contrast, tensor (or gravity) modes
redshift the wavelength of the incident anisotropic radiation along diagonals
to the propagation vector, resulting in a polarization pattern with a non-zero 
curl. The presence of B-mode polarization in the CMB could thus signal a possible 
origin in tensor fluctuations of a quantized scalar field in the early Universe.

But the detection of these modes is quite challenging, because the
foreground polarized intensity due to dust in the Galaxy is orders of
magnitude larger than that expected in the CMB \citep{PlanckVII:2020}.
Aspherical dust particles in our neighborhood align with an ambient 
magnetic field and produce both E-mode and B-mode polarization. 
Unfortunately, a simple modeling of this emission is not possible
because the relative power in the E-mode and B-mode components is a complex 
function of the underlying physical conditions, especially (i) the magnetic 
field {\bf B}, (ii) its geometry (i.e., turbulent versus smooth) and
(iii) its energy density relative to that of the plasma. And there have
been some surprises. The naive prediction had been that the E-mode and
B-mode polarizations would be oriented randomly in the foreground map,
with roughly equal intensities \citep{Caldwell:2017}. {\it Planck} instead
measured a factor $\sim 2$ E/B anisotropy \citep{PlanckVII:2020}.

The signal remaining after this foreground polarization is subtracted
contains only an E-mode pattern that one may assign to the CMB,
with no evidence of any B-mode polarization. An additional significant 
clue is provided by stacking the CMB peaks, which uncovers a 
characteristic ringing pattern in the temperature due to the 
first acoustic peak (on scales of $\lesssim 1^\circ$), with an
associated strong pattern in the E-mode stack (see, e.g., fig.~20 in
\citealt{PlanckVII:2020}). In the context of $\Lambda$CDM, such a 
correlation in the temperature and E-mode anisotropies supports the 
standard picture in which the CMB anisotropies were created in the 
recombination zone. 

As we now discuss, however, neither the TE correlation, nor the absence of 
B-mode polarization in the foreground-subtracted signal, rule out a reprocessing 
of the CMB by a dust screen at $z_{\rm dust}$. We shall demonstrate (i) that
the current observations are not sufficiently precise to do this, and
(ii) that our theoretical understanding of dust polarization is not
complete enough for us to conclude that B-mode polarization should or 
should not be present in the foreground-subtracted CMB map.

For dust to emit polarized light, its grains need to be non-spherical
so that they can spin about their semi-major axis. In addition, there must be an
organized magnetic field present to align them. We have no evidence regarding
whether or not the earliest dust grains produced in Pop III ejecta were spherical,
but observations in other dust environments suggest that they probably were not.
We similarly have no hard evidence for the existence of an intergalactic
magnetic field, but our current measurements suggest that, if present, it 
was weaker than those found in galaxies, where $|{\bf B}_{\rm G}|\sim 3-4\,\mu$G
\citep{Grasso:2001}. 

Nevertheless, it is certainly not ruled out. For example, fields within the 
Abel clusters appear to have amplitudes $|{\rm B}_{\rm ICM}|\sim 1-10\,\mu$G.
Other lines of evidence include high resolution measurements of the rotation 
measure in high-$z$ quasars, which indicate that weak magnetic fields must
have been present in the early Universe. Radio observations of 3C191 at $z=1.945$
\citep{Kronberg:1994} imply that $|{\bf B}_{\rm IGM}|\sim 0.4-4\mu$G.

Interesting limits may also be derived from the ionization fraction in the 
intergalactic medium and reasonable assumptions regarding the magnetic 
coherence length. The largest reversal scale ($\sim 1$ Mpc) seen in galaxy 
clusters implies that $|{\bf B}_{\rm IGM}|\lesssim 10^{-9}$ G 
\citep{Kronberg:1994,Grasso:2001}. If their coherence length is much
larger, however, these fields could be as small as $\sim 10^{-11}$ G. 

This range of values for $|{\bf B}_{\rm IGM}|$ is also supported by
less direct arguments. For example, the galactic dynamo origin for 
${\bf B}_{\rm G}$ is still controversial. Some maintain that the galactic 
field was produced by the adiabatic compression of $|{\bf B}_{\rm IGM}|$
when the protogalactic cloud collapsed. If so, then $|{\bf B}_{\rm IGM}|$
would have been $\sim 10^{-10}$ G at $z>5$, around the time when
galaxies were forming. This argument is consistent with the limits placed 
by the rotation measures of high-$z$ objects \citep{Grasso:2001}.

The magnetic field within the halos where Pop III stars formed and ejected
their dust could have been even stronger than $|{\bf B}_{\rm IGM}|$. The
issue here is whether it was strong enough to align the dust grains. But
other mechanisms have also been proposed to assist with this process,
such as mechanical alignment \citep{Dolginov:1976,Lazarian:1994,Roberge:1995,Hoang:2012}
and radiative alignment \citep{Dolginov:1976,Draine:1996,Draine:1997,Weingartner:2003,Lazarian:2007},
which are not so sensitive to the magnetic field amplitude. It is fair to
say that our experience with dust grain alignment and their polarized emission
within our Galaxy is probably not general enough to encompass such processes
during the Pop III era at $z_{\rm dust}\sim 16$.

This uncertainty is relevant to the question of why polarized dust emission
has never been seen from the intergalactic medium, and whether the CMB
polarization constraints already available today may nevertheless still be
consistent with the dust screen scenario. There are several good reasons
to suspect that this may be the case. First, the dust particles in the
screen at $z_{\rm dust}\sim 16$ would have been destroyed by redshift
$z\sim 14$. As such, the non-detection of polarized dust emission from the 
IGM at $z<14$ does not necessarily imply an absence of a magnetic field
in the intergalactic medium. 

Second, detailed theoretical studies of dust emission and its polarization
(motivated in part by the {\it Planck} data), show that the dust polarization 
fraction is typically $\sim 6-10\%$ for a broad range of physical conditions 
(see, e.g., \citealt{Draine:2009}). This outcome is remarkably similar to the 
$\sim 10\%$ fraction measured in the CMB \citep{PlanckVII:2020}. In other words, 
the dust screen at $z_{\rm dust}\sim 16$ could very well have accounted for all 
of the E-mode polarization fraction measured in the CMB thus far.

The {\it relative} E-mode and B-mode intensities depend on several detailed 
properties of the dust particles and the background magnetic field (see, e.g., 
\citealt{Caldwell:2017,Kritsuk:2018,Kim:2019}). Indeed, an alignment between
the density structures and the magnetic fields generates more E-mode power
than B-mode \citep{Zaldarriaga:2001}. Thus, a third reason why the dust screen
may have produced the observed CMB polarization features is that the E/B 
asymmetry depends heavily on the randomness of this alignment: a higher
degree of randomness produces less E/B asymmetry. If {\bf B} within the 
Pop III halos was highly organized, the E/B asymmetry produced by the
dust would have been quite large.

\cite{Caldwell:2017} considered this feature of the E/B emissivity in
the context of magnetized fluctuations comprised of slow, fast, and 
Alfv\'en magnetohydrodynamic waves. They concluded that the ratio E/B 
could emerge within quite a broad range, from the factor $\sim 2$ observed 
in the Galaxy by {\it Planck}, all the way to $\sim 20$ when the medium 
is threaded by weak fields and fast magnetosonic waves. But these are
precisely the conditions that would have been present in the dust environment 
at $z_{\rm dust}\sim 16$ (see their figure 3 for a summary of these results). 

Thus, contrary to what conventional wisdom would have us believe, polarized 
dust emission need not automatically produce detectable B-mode emission
along with E-mode.  As such, the current non-detection of B-mode polarization 
in the foreground-subtracted CMB intensity does not rule out reprocessing
of the background radiation by dust within the Pop III screen. Interestingly,
an eventual detection of B-mode polarization could work both ways: it could
either constrain inflation in $\Lambda$CDM, or the physical conditions within 
a magnetized dusty environment at $z_{\rm dust}\sim 16$.

Finally, we acknowledge the fact that a TE correlation has also been
detected by {\it Planck} in the foreground dust emission, not just the
direct CMB signal. This means that the overlap between the temperature and 
E-mode stacks of the foreground-subtracted CMB intensity need not necessarily
be due solely to Thomson scattering in a recombination zone. It apparently
could have been produced by polarized dust emission in a dust screen at 
$z_{\rm dust}\sim 16$.

\subsection{Weak lensing of the CMB}
Aside from the observational signatures associated with a dust screen in
the Pop III era that we have discussed thus far, several other lesser known 
characteristics of the CMB can potentially distinguish between an interpretation
of the anisotropies originating at the LSS ($z\sim 1080$) versus $z_{\rm dust}\sim 16$.
One of these is lensing of the CMB, which has been measured to very high precision.
The observed deflections appear to be consistent with the transfer of the background
radiation over a comoving distance stretching from $z\sim 1080$ to $0$ (for
an early review, see \citealt{Lewis:2006}). But this is only true in the context
of $\Lambda$CDM. In the $R_{\rm h}=ct$ universe, one can show that the current
CMB lensing data are instead fully consistent with the analogous transfer of 
background radiation across a comoving distance from $z_{\rm dust}\sim 16$ to $0$.

Although the LSS redshift ($z\sim 1080$) and dust-screen redshift
($z_{\rm dust}\sim 16$) differ considerably in these two scenarios,
what matters most in establishing the lensing effects are: (i) the
comoving distance to the location of the CMB anisotropies, (ii) the
size of the potential wells along the line of sight, and (iii) the
actual pattern of anisotropies where the CMB radiation is released.
Quite critically, the time-redshift relation between these two models
differs by a factor of up to $\sim 2$, which accounts for the majority
of the differences, as we shall see.

The calculations completed thus far for the formation of structure in
the $R_{\rm h}=ct$ universe \citep{Melia:2017a,Yennapureddy:2018a} show 
that the typical potential well in this model had a size of $\sim 265$ Mpc,
compared with $\sim 300$ Mpc in $\Lambda$CDM. The comoving distance
between $z_{\rm dust}\sim 16$ and $0$ is correspondingly $\sim 12,200$ 
Mpc. Thus, the CMB radiation would have traversed approximately $46$ 
potential wells from the dust screen to $z=0$. Ironically,
this is almost exactly the same number as that from $z\sim 1080$ to 
$0$ in the standard model \citep{Lewis:2006}, a similarity that continues
to affirm the suspicion that the various free parameters in $\Lambda$CDM
are optimized by the data to mimic the expansion profile predicted by
the zero active mass condition in $R_{\rm h}=ct$ \citep{Melia:2020}.

We therefore estimate that weak lensing from $z_{\rm dust}\sim 16$ to 
$0$ in $R_{\rm h}=ct$ would have produced an average deflection angle 
very similar to that from $z\sim 1080$ to $0$ in $\Lambda$CDM. Assuming
that the potentials are uncorrelated, we find that the total deflection 
angle should be $\sim\sqrt{46}\times 10^{-4}$ radians, based on the 
deflection angle due to each single well. Quite remarkably, the 
overall deflection angle is therefore $\sim 2$ arcmins in both 
models. Of course, the detailed remapping of the CMB temperature 
profile by weak lensing is much more complicated than this, and ought
to be done in the near future. Since the scales are so similar, however, 
we already know that the observed weak lensing profile probably cannot 
distinguish between an origin of the CMB within a recombination zone 
at $z\sim 1080$ and one due to subsequent reprocessing by a Pop III
generated dust screen in the context of $R_{\rm h}=ct$.
 
\subsection{Growth of Structure}
Finally, we consider whether density fluctuations corresponding to
the CMB temperature anisotropies at $z_{\rm dust}\sim 16$ would have
had enough time to grow into the large-scale structure we observe today.
The fluctuations in the CMB temperature correspond to $\sim 10^{-5}$
amplitude variations in density that are believed to have grown 
gravitationally into galaxies and clusters. In the standard model, 
linear growth between $z\sim 1080$ and today may account for the 
formation of structure beginning with such initial conditions, 
though the early appearance of galaxies and quasars is creating 
some tension with the standard timeline 
\citep{Melia:2013b,Melia:2014a,Yennapureddy:2018a,Yennapureddy:2021}.

But if the CMB anisotropies were instead indicative of features emerging
at $z_{\rm dust}\sim 16$, the standard picture would be entirely unworkable,
given that there would have been barely enough time to accomplish this
starting at $z\sim 1080$. The time-redshift relation in $R_{\rm h}=ct$ 
is sufficiently different, however, to fully compensate for the shorter
range in $z$. With a timeline $t(z) = t_0/(1+z)$ \citep{Melia:2017a}, 
the duration from $z_{\rm dust}\sim 16$ to today is $\sim 13$ Gyr. This
is to be compared with the time elapsed since $z\sim 1080$ in $\Lambda$CDM,
which is $\sim 13.7$ Gyr. These two times are clearly indistinguishable,
which suggests that galaxies and clusters could have grown with equal
viability in the comoving frames of these two scenarios.

\section{Conclusion}\label{conclusion}
We have demonstrated in this paper that the dust model proposed
in the 1970's for the origin of the CMB may still be relevant today,
albeit in the context of the $R_{\rm h}=ct$ cosmology rather than
the standard model, $\Lambda$CDM. Indeed, for a non-inflationary
history, such as that expected in the former, the anisotropies 
measured in the microwave background could not have been established
at the time when protons and electrons combined and liberated the 
CMB relic photons. Instead, all the current observational constraints
point to a model in which the CMB photons were subsequently reprocessed 
by dust at a redshift $z_{\rm dust}\sim 16$. 

This period is interesting but, even more importantly, was followed directly 
by the epoch of reionization, which began at $z_{\rm EoR}\sim 14-15$. The 
alignment of these two redshifts cannot---and should not---be viewed as a 
mere coincidence, because this transition can be easily understood based
on what we see in the local Universe. The rapid formation of stars would 
have filled the Universe with dust during the Pop III era, which presumably 
peaked at $z\sim 16$, followed by a $\sim 100$ Myr period during which 
subsequent supernova explosions would have destroyed all of the dust 
particles. Together with the correlated rapid increase in UV emissivity, 
one expects this transition to have signaled the reionization phase.

We are fortunate in that the observational signatures of the recombination
and dust scenarios should be easily distinguishable with the improved 
sensitivities of future experiments. We should be able to definitly
rule out one or the other of these models over the next few years,
either (i) from the detection of recombination lines at $z\sim 1080$, 
providing compelling evidence in favour of $\Lambda$CDM, or (ii) by confirming
the current hint of frequency dependence in the CMB power spectrum, which
would favour $R_{\rm h}=ct$. A $\sim 5\%$ variation in the anisotropy 
spectrum across the sampled frequency range could be naturally explained 
as a change in the optical depth through the dust screen, but not in 
the Thomson scattering-dominated recombination zone.

\begin{acknowledgments}
I am grateful to Daniel Eisenstein, Anthony Challinor, Martin Rees, Jos\'e Alberto 
Rubino-Martin, Ned Wright, Craig Hogan and Aigen Li for several helpful discussions.
I am also grateful to the anonymous referee for their expert review and
several recommendations that have improved the manuscript.
\end{acknowledgments}

\bibliographystyle{aasjournal}
\bibliography{ms}

\end{document}